\documentclass[runningheads]{svmult}

\usepackage{makeidx}   
\usepackage{graphicx}  
\usepackage{subeqnar}  
\usepackage{multicol}  
\usepackage{physprbb}  
\makeindex             

\begin{document}
\title*{Local characteristic algorithms for relativistic hydrodynamics}
\toctitle{Local characteristic algorithms for relativistic hydrodynamics}
%
%
\titlerunning{Local characteristic algorithms for relativistic hydrodynamics}
%
\author{Jos\'e A. Font}
\authorrunning{Jos\'e A. Font}
%
%
\institute{Max-Planck-Institut f\"ur Astrophysik \\
Karl-Schwarzschild-Str. 1 \\
85740 Garching, Germany}

\maketitle              

\begin{abstract}
 Numerical schemes for the general relativistic hydrodynamic equations
 are discussed. The use of conservative algorithms based upon the
 characteristic structure of those equations, developed during the
 last decade building on ideas first applied in Newtonian hydrodynamics,
 provides a robust methodology to obtain stable and accurate solutions
 even in the presence of discontinuities. The knowledge of the wave
 structure of the above system is essential in the construction
 of the so-called linearized Riemann solvers, a class of numerical schemes
 specifically designed to solve nonlinear hyperbolic systems of
 conservation laws. In the last part of the review some astrophysical 
 applications of such schemes, using the coupled system of the 
 (characteristic) Einstein and hydrodynamic equations, are also 
 briefly presented.
\end{abstract}

\section{Introduction}

The numerical investigation of the Einstein equations is nowadays
an important and fruitful line of research in general relativity. There
exist a number of mathematical formulations of the gravitational
field equations which are on the basis of all numerical approaches.
The level accomplished in the understanding of these equations, both 
mathematically and numerically, is high enough to allow, in principle, 
for sound numerical approaches. Nevertheless, apart from some remarkable 
results, such as the discovery of critical phenomena by 
Choptuik~\cite{choptuik93} or, more recently, the achievement of 
long-term stable three-dimensional null cone evolutions of single 
black hole spacetimes~\cite{gomez98}, the numerical investigations 
have only partially succeeded, quite understandably due to the 
complexity of the theory, in providing global numerical 
solutions of generic spacetimes, especially in the presence 
of curvature singularities. Traditionally, the formulation 
which has received the greatest attention by numerical relativists 
has been the so-called 3+1 (ADM) 
formulation~\cite{lichnerowitz,choquet,arnowitt} (see also the recent review
by Friedrich and Rendall~\cite{friedrich} and references therein). This
formulation of the Einstein equations as a Cauchy (initial value)
problem, along with its multiple variants - hyperbolic (see, 
e.g.~\cite{reula} and references therein) and conformal
reformulations~\cite{shibata,baumgarte} - 
is still today the workhorse of numerical relativity
(for an up-to-date summary of the status of numerical approaches
in 3+1 see, e.g.~\cite{seidel} and references therein.) 

On the other hand, despite being known for about forty years 
now~\cite{bondi,sachs,penrose}, the characteristic formulation
of the Einstein equations has however been used by much less
research groups in numerical relativity (see the
recent review by Winicour~\cite{winicour}). And research programs to 
integrate numerically the conformal equations~\cite{friedrich2} 
have started only more recently~\cite{huebner1,frauendiner,huebner2}.
Compared to the 3+1 formulation, the latter two formulations are best 
suited to study the conformal structure of the spacetime - the main 
topic of the workshop -, and the propagation of radiation fields 
within the spacetime. State-of-the-art numerical methodology applied to
such two formalisms is comprehensively reviewed in the corresponding 
articles by Bartnik and Lehner (characteristic equations) and 
Frauendiener, Husa and Schmidt (conformal equations) in this volume. 
Apart from some test computations involving matter sources presented
in Lehner's article, those papers are mainly concerned with the 
integration of the vacuum field equations.  

The present contribution to this volume aims, on the other hand, 
at describing the current status of the numerical integration of the 
hydrodynamic equations on curved spacetimes, complementing, to some 
extent, the contributions by the above authors. Even though the 
description will be rather basic and general, I will also present 
some applications and some recent results obtained with the 
{\it coupled} integration of the Einstein and hydrodynamic 
equations within the framework of the characteristic formulation of the 
gravitational field equations. It is worth pointing out that, while
initially the characteristic evolution of matter was limited to
idealized systems such as massless scalar fields, nowadays it is
mature enough to account for fully hydrodynamical evolutions with perfect 
fluids~\cite{dubal,bishop,pf2000,pf2001,felix,florian}.

Admittedly, the motivation to develop the capabilities to perform 
coupled evolutions of the matter fields and the geometry needs really not 
much of an emphasis. In astrophysics general relativity plays a major role
in the description of compact objects in such diverse scenarios as
core collapse supernovae, black hole formation, accretion, gamma-ray
bursts and coalescing compact binaries. With the exception of the 
coalescence and merging of two black holes - the number one problem 
of nowadays' numerical relativity - all realistic astrophysical 
systems and sources of (detectable) gravitational radiation involve 
matter. 

The only means to study the time-dependent evolution of fluid 
flow coupled to geometry is through numerical simulations.
Some scenarios can be properly described in the so-called `test fluid' 
approximation, where the self-gravity of the fluid is neglected.
Nowadays there is a large body of numerical investigations in the
literature dealing with such hydrodynamical integrations in {\it static}
background spacetimes (see, e.g., references in~\cite{fontlr}). 
Most of these are based on the pioneering formulation of the hydrodynamic
equations by Wilson~\cite{wilson} and use 
numerical schemes based on finite differences with some amount 
of artificial viscosity. The use of conservative
formulations of the equations, and their characteristic information,
in the design of numerical schemes started in more recent 
years~\cite{mim}.

On the other hand, time-dependent simulations of
self-gravitating flows in general relativity, evolving the spacetime
dynamically with the Einstein equations coupled to a hydrodynamic
source, are more scarce. Although there is much recent interest in 
this direction, only the spherically symmetric case has been 
extensively studied thus far (since the pioneering work of May and
White in 1966~\cite{maywhite}). In axisymmetry, fewer attempts have been 
made, most of them devoted to the study of the gravitational collapse 
of rotating stellar cores and the subsequent emission of gravitational
radiation~\cite{nakamura,stark,dimmelmeier}. The three-dimensional 
efforts are nowadays mainly focused on the study of the dynamics of
relativistic stars~\cite{baumgarte2,font1,alcubierre,stergioulas,font5},
with the detailed study of the coalescence of close neutron star 
binaries being the key target~\cite{mathews,shibata2,uryu,miller}.
These investigations are driven by the emerging possibility of detecting 
gravitational waves in a few years time with the different experimental 
efforts currently underway~\cite{thorne}.

The current article deals with the presentation of the main ideas
concerning a particular kind of the ``specialized techniques" (in the
language of~\cite{isaacson}) used to solve nonlinear hyperbolic systems of
conservation laws with finite differences. The discussion will be
specialized to the general relativistic hydrodynamic equations. These
equations - as well as their limiting counterparts in Minkowski spacetime and
Newtonian gravity - constitute a nonlinear hyperbolic system of conservation 
laws. For such systems there exist ever increasing sound mathematical 
foundations and accurate numerical methodology, imported from Computational 
Fluid Dynamics. The schemes that will be discussed here are the so-called
{\it high-resolution shock-capturing} schemes (HRSC in the following),
based upon Riemann solvers and written in {\it conservation form}.
It is worth noticing that there are a number of excellent textbooks which 
deal with this subject in great detail, in 
particular~\cite{leveque,toro,hirsch,laney}
(see also the contribution by Kreiss in this volume). 
Recent reviews on numerical relativistic hydrodynamics are available as 
well~\cite{ibanez,marti2,fontlr}. The interested reader is addressed to 
these references for a more complete information.

The article is organized as follows: Section 2 presents the relativistic
hydrodynamic equations emphasizing work done on conservative formulations.
Such formulations are well-adapted to the numerical schemes which are
discussed in Section 3. Applications of these algorithms are shown in
Section 4. Finally, Section 5 closes the article with a short summary.

\section{Relativistic hydrodynamic equations}

The general relativistic hydrodynamic equations consist of the local
conservation laws of the stress-energy tensor $T^{\mu \nu}$ (the Bianchi
identities) and of the matter current density $J^{\mu}$ (the continuity
equation):
\begin{equation}
{\nabla}_{\mu} T^{\mu \nu} = 0,
\label{eq:stressenergycons}
\end{equation}
\begin{equation}
{\nabla}_{\mu} J^{\mu} = 0.
\label{eq:masscons}
\end{equation}

As usual ${\nabla}_{\mu}$ stands for the covariant derivative associated with
the four-dimensional spacetime metric, $g_{\mu\nu}$.
The density current is given by $J^{\mu} = \rho u^{\mu}$, where $u^{\mu}$
represents the fluid 4-velocity and $\rho$ is the rest-mass density in a
locally inertial reference frame. Greek (Latin) indices run from 0 to
3 (1 to 3) and geometrized units $G=c=1$ are used in the following.

By neglecting non-adiabatic effects such as viscosity
or heat transfer, the stress-energy tensor of a perfect fluid reads:
\begin{equation}
T^{\mu \nu} = \rho h u^{\mu} u^{\nu} + p g^{\mu \nu},
\label{perf_fluid}
\end{equation}
where $p$ is the pressure and $h$ is the relativistic specific 
enthalpy, defined by
\begin{equation}
h = 1 + \varepsilon + \frac{p}{\rho}.
\end{equation}
The quantity $\varepsilon$ is the specific internal energy.

After choosing an explicit coordinate system $x^{\mu}=(x^{0},x^{i})$ 
the previous conservation equations read:
\begin{eqnarray}
\label{initial1}
\frac{\partial}{\partial x^{\mu}} \sqrt{-g} J^{\mu} & = & 0 \, , \\
\frac{\partial}{\partial x^{\mu}} \sqrt{-g} T^{\mu\nu} & = & - \sqrt{-g}
\Gamma^{\nu}_{\mu\lambda} T^{\mu\lambda} \, ,
\label{initial2}
\end{eqnarray}
where the scalar $x^{0}$ represents a foliation of the spacetime with
a family of hypersurfaces. Furthermore, $\sqrt{-g}$ is the volume element 
associated with the 4-metric, with $g=\det(g_{\mu\nu})$, and
$\Gamma^{\nu}_{\mu\lambda}$ are the 4-dimensional Christoffel symbols.

In addition to the equations of motion~(\ref{eq:stressenergycons}) 
and the continuity equation~(\ref{eq:masscons}) the system must be closed 
with an equation of state (EoS) relating the pressure with some independent 
thermodynamical quantities, such as the rest-mass density and internal energy:
\begin{equation}
p=p(\rho,\varepsilon).
\label{eos}
\end{equation}

Relativistic hydrodynamic flows were first studied numerically with
finite-difference schemes and explicit artificial viscosity
terms~\cite{maywhite,wilson}. These terms~\cite{richtmyer} 
were necessary in order to damp the spurious
numerical oscillations associated with the presence of
discontinuities in the flow solution. Such approaches, 
albeit extensively (and successfully) used in different fields of 
computational relativistic astrophysics (e.g., gravitational collapse, accretion,
coalescence of compact binaries, cosmology), were not able,
however, to simulate flows with Lorentz factors $\gamma$ 
larger than 2~\cite{norman}, for which {\it implicit} methods were
considered more appropriate. More recently, however, the 
use of artificial viscosity terms in non-grid based algorithms such
as Smoothed Particle Hydrodynamics, has proven not to have such 
severe limitations~\cite{siegler}.

The study of ultrarelativistic hydrodynamics with {\it explicit} 
finite-difference methods underwent a revival with the
adoption of conservative formulations of the hydrodynamic equations
and numerical methodology relying upon the hyperbolic nature of such
system. Theoretical advances on the mathematical character of the relativistic
hydrodynamic equations were achieved studying the special relativistic
limit. In Minkowski spacetime, the hyperbolic character of relativistic
(magneto) hydrodynamics was exhaustively studied by Anile and 
collaborators (see~\cite{anile89} and references therein).
The so-called high-resolution Godunov-type schemes, with low
numerical dissipation and oscillation-free representation of discontinuous
solutions, based upon either exact or approximate Riemann solvers were 
extended during the 1990s from classical fluid dynamics to 
relativity~\cite{mim,font0,eulderink,komissarov,banyuls}. Nowadays, there 
exists increasing expertise, both theoretical and numerical, to 
investigate extremely fast flows through accurate computer simulations 
(see~\cite{ibanez} for a recent review).

Traditionally, most of the approaches for numerical integrations of the
general relativistic hydrodynamic equations have adopted spacelike foliations 
of the spacetime, within the 3+1 formulation~\cite{wilson,banyuls,font1}
Covariant and conservative formulations for ideal fluids, 
have been presented in~\cite{eulderink} and~\cite{pf2000}. 
From the theoretical point of view most of the
existing formulations of the relativistic hydrodynamics 
equations are written in terms of quantities measured by an 
Eulerian (fixed) observer.  By using Eulerian frame variables 
(relativistic densities of mass, momentum and energy) the equations 
exhibit a conservation form similar to their nonrelativistic 
counterparts. In most cases, contrary to Newtonian
hydrodynamics, fulfilling the (desirable) conservation
properties is accompanied by a nonlinear recovery process to
extract physical (primitive) quantities (such as rest-mass density,
sound speed, etc) from the conserved quantities forming
the state vector of the system~\cite{eulderink,banyuls,pf2000}
(we note that a different approach based upon a {\it primitive-variable}
formulation is given in~\cite{wen,komissarov}).

As an example, in the formulation developed by Papadopoulos and 
Font~\cite{pf2000} the spatial velocity components of the 4-velocity, 
$u^{i}$, together with the rest-frame density and internal energy,
$\rho$ and $\varepsilon$, are taken as the primitive variables. They
constitute a vector in a five dimensional space $ {\bf w} = (\rho,
u^{i}, \varepsilon)$. The initial value problem for
equations~(\ref{initial1}) and~(\ref{initial2}) is defined in terms
of another vector in the same fluid state space, namely the {\em conserved
variables}, ${\bf U}=(D,S^{i},E)$:
\begin{eqnarray}
\label{eq:D}
D     & \equiv & J^{0}  = \rho u^{0} \, ,  \\
S^{i} & \equiv & T^{0i} = \rho h u^{0} u^{i} + p g^{0i} \, ,  \\
E     & \equiv & T^{00} = \rho h u^{0} u^{0} + p g^{00} \, .
\label{eq:E}
\end{eqnarray}
With those definitions the hydrodynamic equations can be written as a
first-order flux-conservative hyperbolic system of conservation laws:
\begin{equation}
\frac{\partial(\sqrt{-g} {\bf U})}{\partial x^0}
+ \frac{\partial(\sqrt{-g} {\bf F}^{j})}{\partial x^j}
= {\bf S} \, .
\label{eq:cons-law}
\end{equation}
The flux vectors ${\bf F}^{j}$ and the source terms ${\bf S}$ are given by:
\begin{eqnarray}
{\bf F}^{j} = (J^{j}, T^{ji}, T^{j0}) =
(\rho u^{j},  \rho h u^{i} u^{j} + p g^{ij},
\rho h u^{0} u^{j} + p g^{0j})
\label{eq:fluxes}
\end{eqnarray}
\begin{eqnarray}
{\bf S} = (0, - \sqrt{-g} \, \Gamma^{i}_{\mu\lambda} T^{\mu\lambda},
- \sqrt{-g} \, \Gamma^{0}_{\mu\lambda} T^{\mu\lambda}).
\label{pf:sources}
\end{eqnarray}

The local characteristic structure of these equations has been
presented in~\cite{pf2000}. For the other conservative formulations
mentioned above such information can be found 
in Refs.~\cite{eulderink,banyuls,font1} (see also~\cite{ibanez2}). The
relevance of having the wave structure to one's disposal in the
development of HRSC schemes will become apparent in the following
section.

\section{High-resolution numerical schemes}

The hydrodynamic equations constitute a nonlinear hyperbolic system
of conservation laws. Hence, smooth initial data can turn into discontinuous 
data (i.e., crossing of characteristics in the case of shocks) after a finite 
time during the evolution. Standard finite difference algorithms 
suffer from important deficiencies when dealing with such systems.  
Typically, first order accurate schemes are too dissipative across 
discontinuities (excessive smearing) while second order (or higher) 
schemes produce spurious oscillations near discontinuities, which do 
not disappear as the grid is refined. 

Finite difference numerical schemes provide solutions of the discretized 
version of the original system of partial differential equations. Therefore, 
convergence properties under grid refinement must be enforced on such schemes 
to ensure the correctness of the numerical result (i.e., the global error of 
the numerical solution must vanish as the cell width is diminished). For 
hyperbolic systems of conservation laws, schemes written in {\it conservation
form} are preferred, since - as proven by Lax and Wendroff~\cite{lax} - if 
convergence exists, it is to one of the so-called
{\it weak solutions} of the original system of equations.
Such weak solutions are generalized solutions
that satisfy the integral form of the conservation system $\partial_t{\bf U} +
\partial_x{\bf F}=0$:
\begin{eqnarray}
\int_0^{\infty}\int_{-\infty}^{+\infty}
(\Phi_t {\bf U} + \Phi_x {\bf F}({\bf U})) dx dt =
-\int_{-\infty}^{+\infty} \Phi(x,0){\bf U}(x,0) dx,
\label{weak}
\end{eqnarray}
for any continuously differentiable test function 
$\Phi(x,t)$ with compact support.
They are classical solutions (continuous and differentiable)
in regions where they are smooth and have a finite number of discontinuities.

The class of all weak solutions is too wide in the sense that there
is no uniqueness for the initial value problem. The numerical method should
guarantee convergence to the {\it physically admissible solution}. This
is the vanishing-viscosity solution of the ``viscous version" of the
hyperbolic problem:
\begin{equation}
\frac{\partial {\bf U}}{\partial t} + \frac{\partial {\bf F}({\bf U})}{\partial x} =
\eta\frac{\partial^2{\bf U}}{\partial x^2},
\end{equation}
when $\eta\rightarrow 0$. Mathematically, this solution is characterized by 
the so-called {\it entropy condition} (e.g., the entropy of any fluid
element should increase when running into a discontinuity). The characterization
of the {\it entropy-satisfying solutions} for hyperbolic systems of conservation 
laws was developed by Lax~\cite{lax72}.

The Lax-Wendroff theorem~\cite{lax}  does not establish whether the
method converges, for which some form of {\it stability} is required.
Building upon the {\it Lax equivalence theorem} (see, e.g.~\cite{richtmyer67}),
the notion of total-variation stability has proven very successful, 
although sound results have only been obtained for (nonlinear) 
scalar conservation laws. The total variation of a numerical
solution at time $t=t^n$, TV$({\bf u}^n)$, is defined as:
\begin{equation}
\mbox{TV}({\bf u}^n)=
\displaystyle\sum_{i=0}^{+\infty} |{\bf u}_{i+1}^n-{\bf u}_i^n|.
\end{equation}

A numerical scheme is TV-stable if TV$({\bf u}^n)$ is bounded 
at any time and for any initial data. Present-day research is
focused on the development of high-resolution numerical schemes in
conservation form satisfying the condition of TV-stability, such as
the so-called {\it Total Variation Diminishing} (TVD)
schemes~\cite{harten84}. An additional property
that any numerical method should satisfy is monotonicity (see, 
e.g.~\cite{leveque}). For scalar conservation laws it has been shown that
monotone methods are TVD and satisfy a discrete entropy condition.
Therefore, they converge in a non-oscillatory manner to the unique
entropy (physical) solution.

In a conservative scheme the time variation of the mean values of the 
state vector ${\bf U}$ in a given numerical cell - labelled by index $i$ - 
is given, in the absence of source terms, by the flux differences across
the cell interfaces. Mathematically, such an algorithm reads:
\begin{eqnarray}
{\bf U}_i^{n+1}={\bf U}_i^n - \frac{\Delta t}{\Delta x}
(\hat{\bf F}_{i+\frac{1}{2}} - \hat{\bf F}_{i-\frac{1}{2}}),
\label{conservation}
\end{eqnarray}
where $\Delta t$ and $\Delta x$ stand for the time step and cell width,
respectively.

Historically, in 1959 Godunov~\cite{godunov} developed the first conservative
scheme for the classical fluid equations in which the {\it numerical fluxes},
$\hat{\bf F}_{i+\frac{1}{2}}$, at every cell interface of the computational
grid were computed by exactly solving a family of local
{\it Riemann problems}.
Such Riemann problems - the simplest initial value problem with discontinuous
initial data - arise naturally after the discretization procedure of the
``continuous" solution by means of piecewise constant approximations. 
The Riemann problem is invariant under similarity transformations
$(x,t)\rightarrow (ax,at), a>0$. The solution is therefore constant
along the characteristics $x/t=const.$ and, hence, self-similar.
It consists of constant states separated by rarefaction waves, shocks
and contact discontinuities.

Given a general hyperbolic system, 
if ${\bf U}(x,t)={\bf w}_{R}(x/t;{\bf U}_-,{\bf U}_+)$
is the weak solution of a Riemann problem with initial data ${\bf U}={\bf U}_-$
if $x<0$ and ${\bf U}={\bf U}_+$ otherwise, then the numerical fluxes in
Godunov's scheme are given by:
\begin{eqnarray}
\hat{\bf F}_{i+\frac{1}{2}} = 
{\bf F}({\bf w}_{R}(0;{\bf U}_i^n,{\bf U}_{i+1}^n)),
\end{eqnarray}
along the characteristic $x/t=0$.

The exact solution of the Riemann problem was extended to
relativistic hydrodynamics in~\cite{marti3} (see also~\cite{pons1}).
Since it involves solving a nonlinear algebraic
system which can be computationally inefficient, in addition to the
approximation involved in using piecewise constant data, the use
of approximate solutions of the Riemann problem were proposed.
Hence, if ${\bf w}(x/t;{\bf U}_-,{\bf U}_+)$ is such an approximation,
the Godunov-type schemes are defined~\cite{harten83,einfeldt} as those
in which ${\bf U}_{i}^{n+1}$ are computed as:
\begin{eqnarray}
{\bf U}_{i}^{n+1}&=&\frac{1}{\Delta x}
\displaystyle\int_{-\Delta x/2}^0 {\bf w}(x/t;{\bf U}_i,{\bf U}_{i+1})dx +
\nonumber \\ & &
\frac{1}{\Delta x}
\displaystyle\int_0^{\Delta x/2} {\bf w}(x/t;{\bf U}_{i-1},{\bf U}_{i})dx,
\end{eqnarray}
and the numerical fluxes in the spacetime computational cell
$[x_{i-\frac{1}{2}},x_{i+\frac{1}{2}}]\times[t^n,t^{n+1}]$ are given by:
\begin{eqnarray}
\hat{\bf F}_{i+\frac{1}{2}} = \frac{1}{\Delta t}
\displaystyle\int_{t^n}^{t^{n+1}} {\bf F}({\bf U}(x_{i+\frac{1}{2}},t)) dt,
\end{eqnarray}
where ${\bf U}(x_{i+\frac{1}{2}},t)$ is computed by (approximately)
solving a Riemann problem at every numerical cell interface
${\bf U}(x_{i+\frac{1}{2}},t)={\bf w}(0;U_i^n,{\bf U}_{i+1}^n)$.

The mathematical and algorithmical developments accomplished in the
scalar case have been extended to nonlinear hyperbolic systems of
conservation laws using the so-called {\it local characteristic approach}. 
This technique generalizes the original procedure due to Roe~\cite{roe81}
by applying the scalar algorithms to any of the characteristic
equations of the system, after a suitable linearization. At each interface
$i+1/2$ of the computational grid, the Jacobian matrix
${\bf A}$ of the system, ${\bf A}=\partial {\bf F}/\partial{\bf U}$
is assumed to be constant $\widetilde{\bf A}_{i+1/2}={\bf A}({\bf U}_{i+1/2})$,
with ${\bf U}_{i+1/2}$ being an average between ${\bf U}_i$ and
${\bf U}_{i+1}$.
The original nonlinear system is then rewritten as $\partial_t{\bf U} +
\widetilde{\bf A}\partial_x{\bf U}=0$. The eigenvalues of this matrix are
the characteristic speeds of the Riemann problem. The approximate Riemann
solver obtains the exact solution of the linearized system, which can
be easily computed by solving a system of decoupled, linear characteristic
(scalar) equations. The properties that the matrix $\widetilde{\bf A}$ has
to fulfill can be found in~\cite{roe81} for the widely used Roe's
approximate Riemann solver.

As an illustrative example, for a second-order upwind,
monotone scheme such as MUSCL~\cite{vanleer}, the expression for the
numerical flux function reads:
\begin{eqnarray}
\hat{\bf F}_{i+\frac{1}{2}} = \frac{1}{2}
\left(
{\bf F}({\bf U}^{\rm R}_i) +
{\bf F}({\bf U}^{\rm L}_{i+1}) -
\sum_{n=1}^p
| \widetilde{\lambda}^{(n)}_{i+\frac{1}{2}} |
\Delta \widetilde{{\bf w}}^{(n)}_{i+\frac{1}{2}}
\widetilde{\bf r}^{(n)}_{i+\frac{1}{2}} \right).
\label{nflux}
\end{eqnarray}
Index $p$ indicates the dimensions of the system.
The quantities ${\bf w}={\bf R}^{-1}{\bf U}$
are the so-called {\it characteristic variables},
${\bf R}$ being the matrix whose columns are the
right-eigenvector expressions
of the Jacobian matrix associated with the vector of fluxes.
Furthermore, $\lambda$ and ${\bf r}$ stand for the
eigenvalues and right-eigenvectors of such Jacobian matrix.
The ``tilde" indicates that all quantities have to be
computed with respect to the linearized Jacobian
matrix $\widetilde{\bf A}$.

The jumps of the characteristic variables at each cell interface
are obtained by projecting the jumps of the state-vector
variables with the left-eigenvectors matrix:
\begin{equation}
\Delta \widetilde{{\bf w}}_{i+\frac{1}{2}}
= \widetilde{\bf R}^{-1}_{i+\frac{1}{2}}
({\bf U}^{\rm L}_{i+1} - {\bf U}^{\rm R}_{i}).
\end{equation}

The left (L) and right (R) states of the conserved quantities
${\bf U}$ - at any cell interface - are computed from
the cell-centered values after a suitable monotone
{\it reconstruction procedure}. The way those
variables are obtained determines the spatial order of the numerical
algorithm and controls, in turn, the local jumps at every interface.
A wide variety of cell reconstruction procedures is available in the
literature (see, e.g.~\cite{vanleer,colella,harten87}).

The last term in the flux-formula, Eq.~(\ref{nflux}), represents
the ``numerical viscosity" of the conservative scheme.
The wave structure of the system is thus used to provide the smallest
amount of numerical dissipation yielding accurate solutions of
discontinuities without excessive smearing, avoiding, at the same
time, the growth of spurious numerical oscillations.

So far we have only considered one-dimensional systems of conservation 
laws. For multidimensional
hyperbolic systems containing source terms a standard procedure to
apply the above schemes is to use dimensional splitting - computing the
numerical fluxes along every spatial direction independently -, possibly 
in combination with a method of lines. Therefore, for a three-dimensional 
hyperbolic system:
\begin{eqnarray}
        \frac{\partial \bf U \rm}{\partial t}+
        \frac{\partial \bf F(U) \rm}{\partial x}+
        \frac{\partial \bf G(U) \rm}{\partial y}+
        \frac{\partial \bf H(U) \rm}{\partial z}=
        \bf S(U) \rm,
\end{eqnarray}
where ${\bf F}, {\bf G}$ and ${\bf H}$ are the fluxes in the $x$, $y$ and
$z$ directions, respectively, the dimensional splitting algorithm reads:
\begin{eqnarray}
{\bf U}^{n+1}_{i,j,k} = {\cal L}_s^{\Delta t/2}
                        {\cal L}_h^{\Delta t}
                        {\cal L}_g^{\Delta t}
                        {\cal L}_f^{\Delta t}
                        {\cal L}_s^{\Delta t/2}{\bf U}^{n}_{i,j,k},
\end{eqnarray}
where ${\cal L}_f$, ${\cal L}_g$ and ${\cal L}_h$ denote the operators
associated with the corresponding one-dimensional PDEs, i.e. the operators 
computing the
numerical fluxes at every cell interface in a given direction. Furthermore,
${\cal L}_s$ is the operator which solves a system of ODEs for the
source terms:
\begin{eqnarray}
\frac{\partial \bf U \rm}{\partial t} = \bf S(U) \rm.
\end{eqnarray}
The state vector ${\bf U}$ at the final time $t^{n+1}$ is then computed 
in consecutive sub-steps. On the other hand, in the method of lines 
the time update of all directions - and of the source terms - is done 
simultaneously. The conservative algorithm reads:
\begin{eqnarray}
  \frac{d{\bf U}_{i,j,k}(t) }{d t} & = &
  -\frac{\widehat{\bf F}_{i+{1\over 2},j,k}- \widehat{\bf F}_{i-{1\over 2},j,k}}
  {\Delta x} -  \nonumber \\
                                   &   &
  -\frac{\widehat{\bf G}_{i,j+{1\over 2},k}- \widehat{\bf G}_{i,j-{1\over 2},k}}
  {\Delta y} - \nonumber \\
                                   &   &
  -\frac{\widehat{\bf H}_{i,j,k+{1\over 2}}- \widehat{\bf H}_{i,j,k-{1\over 2}}}
  {\Delta z} + \hat{\bf S}_{i,j,k},
\end{eqnarray}
where the numerical fluxes are given by
  \begin{eqnarray}
  \hat{\bf F}_{i+{1\over 2},j,k} &=& {\bf F}({\bf U}_{i-q,j,k},
  {\bf U}_{i-q+1,j,k},...,{\bf U}_{i+q,j,k}),
\nonumber \\
  \hat{\bf G}_{i,j+{1\over 2},k} &=& {\bf G}({\bf U}_{i,j-q,k},
  {\bf U}_{i,j-q+1,k},...,{\bf U}_{i,j+q,k}),
\nonumber \\
  \hat{\bf H}_{i,j,k+{1\over 2}} &=& {\bf H}({\bf U}_{i,j,k-q},
  {\bf U}_{i,j,k-q+1},...,{\bf U}_{i,j,k+q}),
  \end{eqnarray}
$q$ indicating the stencil chosen to compute these fluxes.
Further details about multidimensional systems and source terms,
with particular emphasis in numerical schemes for {\it stiff} source
terms, can be found in~\cite{leveque2}.

We end this section by pointing out that during the last few years
most of the {\it classical} approximate Riemann solvers
developed in fluid dynamics have successfully been extended to
relativistic hydrodynamics. The interested reader is referred
to~\cite{marti2} for a comprehensive description of such solvers
in relativistic hydrodynamics.

\section{Applications}

We now present some applications of the concepts introduced in the
previous section. In particular we will show some results concerning the
numerical evolution of the equations of hydrodynamics and the gravitational
field within the context of the characteristic formulation of General
Relativity. But let us start first with a demonstration in Minkowski
spacetime.

\subsection{Shock tube test}

A standard test to calibrate a hydrodynamics code based on
the schemes discussed in the previous section
is the so-called shock tube problem. This is a particular
version of a Riemann problem in which the initial states at
both sides of a discontinuity are at rest. Therefore, the state
of the fluid at either side of the interface only differs in
its thermodynamic quantities such as the density and the
pressure.

When the interface is removed, the fluid
evolves in such a way that four constant states develop. In between each state 
there can exist one of three elementary waves: a shock wave, a contact
discontinuity and a rarefaction wave. As mentioned in the preceding
section the exact wave pattern of the (ideal) fluid
state at any given time was first obtained by Godunov~\cite{godunov} in
Newtonian hydrodynamics. Its generalization to relativistic hydrodynamics
was accomplished by~\cite{marti3} (see also~\cite{pons1}). This time-dependent 
problem provides a simple test of the shock-capturing properties of any
numerical scheme, its level of difficulty depending on the initial data.
A comprehensive survey of the behavior of a large
sample of schemes applied to the shock tube problem is presented 
in~\cite{marti2} (the interested reader is also referred to~\cite{marti2}
for further information on tests commonly used to validate numerical 
schemes for the hydrodynamic equations).

The main differences between the solution of relativistic shock tubes 
and their Newtonian counterparts are due to the nonlinear addition of 
velocities and to the Lorentz contraction. The first effect yields a curved 
profile for the rarefaction fan, as opposed to a linear one in the 
Newtonian case. The Lorentz contraction narrows the shock plateau. 
These effects, especially the latter, become particularly noticeable in 
the ultrarelativistic regime ($\gamma\gg 1$).

\begin{figure}[t]
\begin{center}
\includegraphics[width=.77\textwidth,height=0.70\textwidth]{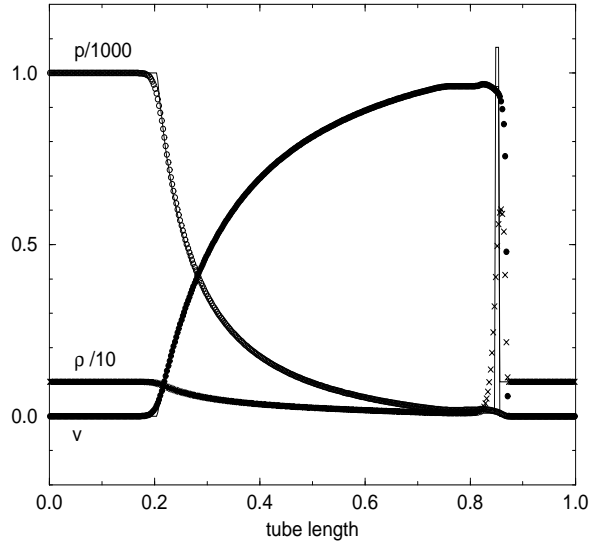}
\end{center}
\caption{
The relativistic shock tube problem at time $t=0.35$. Normalized
profiles of density, pressure and velocity vs. distance for the computed
(symbols) and exact (solid line) solution. All variables were
calculated with a third order scheme on an equidistant grid of 400 zones.
The initial interface was located at $x=0.5$.}
\label{tube}
\end{figure}

For our demonstration we consider a fluid whose initial state is
specified by $p_L=10^3$ and $\rho_L=1$ on the left side of the
interface and by $p_R=10^{-2}$ and $\rho_R=1$ on the right side.
This corresponds to problem 2 of~\cite{marti2}.
The adiabatic index of the perfect fluid EoS, $p=(\Gamma-1)\rho
\varepsilon$, is $\Gamma=5/3$.
An initial jump in pressure of five orders of magnitude leads to the
formation of a thin and dense shell bounded by a leading
shock front and a trailing contact discontinuity - a blast wave. 
The post-shock velocity is 0.96$c$ (Lorentz factor
$\gamma \approx 3.5$), while the shock
speed is 0.986$c$ ($\gamma \approx 6$).
Resolving the thin shock plateau poses a
challenge for any numerical scheme.

Fig.~\ref{tube} shows the results of the shock tube evolution employing a grid
of 400 zones spanning a domain of unit length. The time of the comparison 
between the numerical and the analytic solution is $t=0.35$. We have used 
a HRSC scheme based on the HLLE Riemann solver~\cite{harten83,einfeldt} and 
a parabolic reconstruction procedure~\cite{colella}.
The solid lines indicate the exact solution. Correspondingly, the symbols 
represent the numerical approximation for the (scaled) pressure (open
circles), density (cross signs) and velocity (filled circles).
As one can clearly see the location and propagation speeds of the
different features of the solution are accurately captured. The shock
plateau can be better resolved by simply increasing the numerical 
resolution (see~\cite{donat,marti2}). Diffusion-free results obtained
with a one-dimensional exact Riemann solver are presented in~\cite{wen}.

\subsection{Gravitational collapse of supermassive stars}

Supermassive black holes (SMBH), with masses on the range $10^6 M_{\odot}$-
$10^9 M_{\odot}$ ($M_{\odot}$ indicating the mass of the Sun) are commonly 
found in the center of galaxies~\cite{rees,kormendy}.
Supermassive stars (SMS) have been proposed as possible progenitors of SMBH.
Such stars can develop a dynamical instability~\cite{chandra,fowler}
and undergo catastrophic gravitational collapse.

Recently, the gravitational collapse of SMS was proposed by Fuller 
and Shi~\cite{fuller} as a possible model for gamma-ray bursts. The 
neutrino emission from the collapse
of a SMS could lead to energy deposition by $\nu\overline{\nu}$-annihilation
$\nu_{e,(\mu,\tau)} + \overline{\nu}_{e,(\mu,\tau)}\rightarrow e^{-} + e^{+}$.
Subsequently, $\gamma$-radiation would be
produced by cyclotron radiation and/or the inverse Compton process.

Trying to shed some light on the viability of that mechanism, Linke et
al~\cite{felix} have studied numerically the gravitational collapse of 
spherical SMS using a general relativistic hydrodynamics code. The
code is based on
the hydrodynamics formulation developed by~\cite{pf2000}. The coupled 
system of Einstein and fluid equations is solved adopting a spacetime
foliation with outgoing null hypersurfaces. In such framework and in 
spherical symmetry, the Bondi-Sachs metric~\cite{bondi,sachs} reads:
\begin{equation}
  \label{eq:BondiSachs}
  \mathrm{d}s^{2}= - \frac{e^{2 \beta}V}{r}\mathrm{d}u^{2}-2
  e^{2 \beta}\mathrm{d}u \mathrm{d}r +
  r^{2}(\mathrm{d}\theta^{2} +\sin^{2}{\theta}\mathrm{d}\phi^{2}).
\end{equation}
The Einstein equations simply reduce to two radial hypersurface
equations (ODEs) for $\beta(u,r)$ and $V(u,r)$:
\begin{eqnarray}
  \label{eq:MetricODEConsbeta}
  \beta_{,r}&=&2\pi r e^{4 \beta} E, \\
  \label{eq:MetricODEConsV}
  V_{,r}&=& e^{2 \beta} - 8 \pi r^{2} e^{4\beta}S^{r} - 4 \pi r
  e^{4\beta}V E,
\end{eqnarray}
where ``," indicates partial differentiation.
Correspondingly, the hydrodynamic equations are given by 
expressions~(\ref{eq:D})-(\ref{pf:sources}), appropriately 
particularized to spherical symmetry, and are solved using HRSC
schemes~\footnote{
The reader must be aware of the different meaning of the
word ``characteristic" in the context of the Einstein equations and the
hydrodynamic equations: while in the former case the {\it characteristic 
formulation} of general relativity refers to a particular slicing of the 
spacetime - by means of null cones - in the latter case the notion of the 
{\it local characteristic approach} refers to a numerical procedure seeking
to exploit, algorithmically, the upwind character of the hydrodynamic 
equations.}.

\begin{figure}[t]
\begin{center}
\includegraphics[width=.97\textwidth,height=1.00\textwidth]{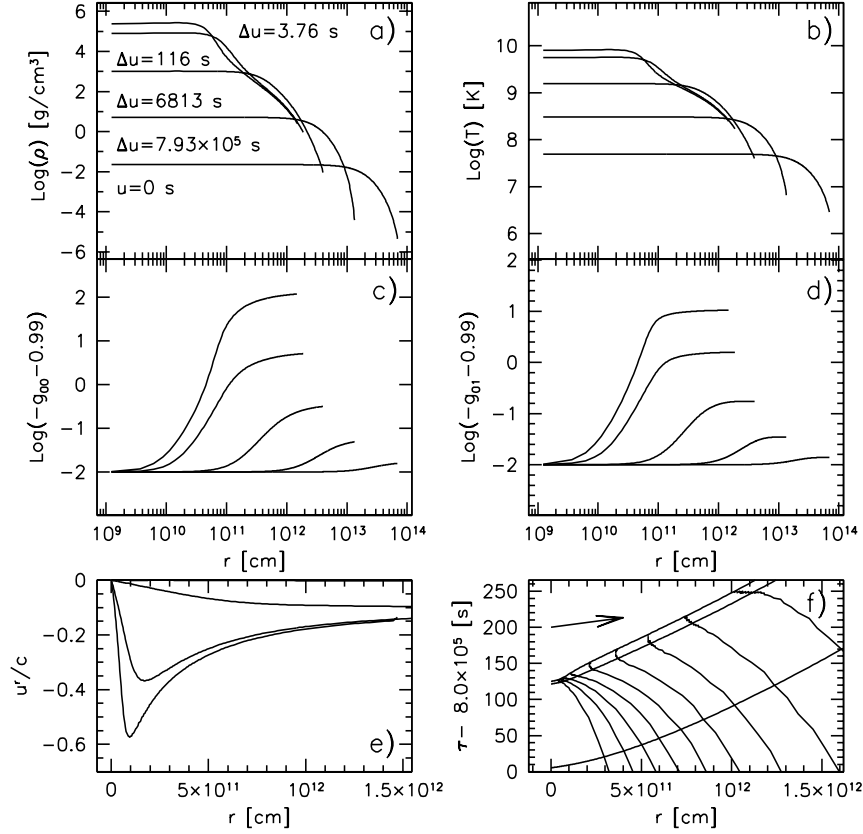}
\end{center}
\caption{Gravitational collapse of a $5\times10^{5} M_{\odot}$ SMS. 
Snapshots of the evolution of the density, temperature, metric
components $g_{uu}$, $g_{ur}$ and radial velocity $u^{r}$.
The spacetime diagram of the lower right panel shows the
location of mass shells ($\Delta
M=5\times10^{4}M_{\odot}$) versus local proper time $\tau$. The
lines intersecting the mass shells are hypersurfaces of
constant coordinate time $u$ and represent trajectories of
outgoing lightrays. A black hole forms enclosing $\sim 25\%$
of the mass of the SMS. The figure is taken from~\cite{felix}.}
\label{SMS1}
\end{figure}

In addition, the code developed by~\cite{felix} includes a tabulated 
EoS which accounts for contributions from radiation, electron-positron pairs 
and baryonic gases, as well as energy losses by thermal neutrino emission.

A typical simulation of a collapsing SMS is depicted in 
Figure~\ref{SMS1}. The initial model corresponds to a $5\times10^{5} 
M_{\odot}$ SMS. The figure shows the radial profiles of the evolution of 
the density, temperature, metric components $g_{uu}$, $g_{ur}$, and
radial velocity. Furthermore, the spacetime diagram at the
lower right panel shows the local proper time against the location 
of mass shells enclosing fixed fractions of the total mass of the star.
The arrow indicates the slope of a lightray in Minkowski spacetime.
One can see that lightrays are severely delayed close to the forming
black hole. This black hole forms from the innermost 25\% of the
total stellar mass. The collapse lasts $8\times 10^5$s ($\sim 9.3$ days) 
and the central density increases by a factor of $1.08\times 10^7$. 
The final configuration becomes highly relativistic before the 
simulation is stopped, with $g_{uu}=-119$ at the surface of the star
($g_{uu}=-1.0058$ initially). Details about the neutrino emission 
in such an evolution can be found in~\cite{felix}.

This and other simulations have been used in~\cite{felix} to analyze the 
possibility that collapsing SMS could be progenitors of gamma-ray 
bursts~\cite{fuller}. The comprehensive study performed 
in~\cite{felix} reveals that
$99\%$ of the energy produced by $\nu\overline{\nu}-$annihilation
is deposited in a spherical layer deep inside the star at a radius
$R_{\nu}\le r \le 3R_{\nu}$, where $R_{\nu}$ indicates the location 
of the neutrino radiating volume. Therefore, only a tiny fraction 
of the energy is deposited near the surface of the star where 
excessive baryon loading could be avoided. As a result, 
ultrarelativistic ejection of matter with Lorentz factors 
$\gamma\gg 1$, a distinctive feature of all gamma-ray burst models,  
cannot be expected in spherical models.
The simulations performed in~\cite{felix} show that 
the spherical collapse of a SMS ($M\ge 5\times 10^5 M_{\odot}$) 
does not meet the demands for being a successful central engine 
for a gamma-ray burst.

\subsection{Null cone evolution of relativistic stars}

In~\cite{florian} we presented the first results of a program we have
recently started to study the dynamics of relativistic stars by
means of null cone simulations in axisymmetry. The final aim is to
study the gravitational core collapse problem and to compute the
associated gravitational radiation~\cite{zwerger,dimmelmeier}.

For these investigations we use the Bondi-Sachs metric:
\begin{eqnarray}
\nonumber
ds^{2} & = & -\left(\frac{V}{r}e^{2 \beta} -
U^{2} r^{2} e^{2 \gamma}\right) du^{2} - 2 e^{2 \beta} du \
dr
- 2 U r^{2} e^{2 \gamma} du \ d\theta \\
\label{Bondi}
&  & + r^{2} (e^{2 \gamma} d
\theta^{2} + e^{-2 \gamma} sin^{2}\theta \ d \phi^{2}),
\end{eqnarray}
with null coordinate $u$, radial coordinate $r$, polar coordinate
$\theta$ and the azimuthal coordinate $\phi$, which is a Killing
coordinate. Using this metric the Einstein equations split into
hypersurface equations on each light cone (for the fields $\beta$, $U$
and $V$), and one evolution equation (for the field $\gamma$, 
not to be confused with the Lorentz factor of the fluid), a wave equation
(see also Lehner's article in this volume). The hypersurface equations,
$G_{1\nu} - 8\pi T_{1\nu}=0$, read:
\begin{eqnarray}
\beta_{,r} &=& \frac{1}{2} r\: (\gamma_{,r})^2
-\frac{1}{4}r R_{rr},
\end{eqnarray}
\begin{eqnarray}
 [r^4 \, e^{2(\gamma-\beta)} U_{,r}]_{,r} &=&
  2 r^2 \left[ r^2  \left(\frac{\beta}{r^{2}}\right)_{,r\theta}
 - \frac{(\sin^2 \theta \,\gamma)_{,r\theta}}{\sin^{2}\theta}
 + 2\,\gamma_{,r} \, \gamma_{,\theta}) \right]
\nonumber \\
&& -2r^2 R_{r\theta},
\end{eqnarray}
\begin{eqnarray}
 V_{,r}  &=& -\frac{1}{4} r^4 e^{2(\gamma-\beta)}(U_{,r})^2
  + \frac{(r^4 \sin\theta U )_{,r\theta}}{2 r^2 \sin\theta}
  \nonumber \\
& & +e^{2 (\beta - \gamma)}\left[
 1 - \frac{(\sin \theta \beta_{,\theta})_{,\theta}}{\sin\theta}
 + \gamma _{,\theta\theta} + 3 \cot\theta \gamma_{,\theta}
 - (\beta_{,\theta})^2
 \right. \nonumber \\ & & \left.
 -2 \gamma_{,\theta} (\gamma_{,\theta} -\beta_{,\theta})
 -\frac{1}{2}r^2 e^{2\beta} g^{AB}R_{AB}
 \right],
\end{eqnarray}
where $R_{\mu\nu}$ is the Ricci tensor and $x^A=(\theta,\phi)$.
Correspondingly, the evolution equation for the gravitational field
reads:
\begin{eqnarray}
4 r ( r \gamma)_{,ur}  & = &  \left[ 2 r \gamma_{,r} V
- r^{2} \left( 2 \gamma_{,\theta}\,U
+ \sin \theta \left({U \over {\sin \theta}}\right)_{,\theta} \right)
\right]_{,r}
\nonumber \\
& & -2 r^{2} \frac{(\gamma_{,r} U \sin\theta)_{,\theta}}{\sin\theta}
+ \frac{1}{2} r^{4} e^{2(\gamma-\beta)} (U_{,r})^{2}
\nonumber \\
& & + 2 e^{2(\beta - \gamma)} \left[ (\beta_{,\theta})^2
+ \sin \theta \left({\beta_{,\theta} \over {\sin \theta}}\right)_{,\theta} 
\right] 4\pi (\rho +p) u_{\theta}^2.
\label{eq:gammaev}
\end{eqnarray}

A remarkable property of the above system of equations is that
they form a hierarchy: knowing $\gamma$ on the first null
hypersurface allows one to radially integrate the corresponding
equations to determine $\beta$, $U$, $V$ and $\gamma_{,u}$
(in that order) on that hypersurface~\cite{winicour}.

In the code developed by~\cite{florian} the numerical implementation
of the Einstein equations closely follows that of~\cite{gomez84}.
The same marching algorithms are employed with additional source
terms arising from the presence of matter fields. Since the code
uses spherical coordinates, special care is taken with the numerical
treatment of the coordinate singularities at the origin and at the
polar axis. In order to impose boundary conditions at the origin the
assumption that $t=u+r$, $x=r\sin\theta\cos\phi$, $y=r\sin\theta\sin\phi$
and $z=r\cos\theta$ form a local Fermi system at $r=0$ is enforced.
This implies a fall-off behavior given by $V=r+O(r^3)$, $\beta=O(r^2)$,
$U=O(r)$ and $\gamma=O(r^2)$~\cite{isaacson}. Regularity on the 
axis requires that
$U/\sin\theta$ y $\gamma/\sin^2\theta$ are continuous functions at
$\theta=0, \pi$. The code also uses a new polar coordinate
$y=-\cos\theta$. In order to keep the freedom of working with
numerical grids which only cover the star without its vacuum
exterior, the radial coordinate used in~\cite{gomez84} was
generalized: Starting from an equidistant radial coordinate
$x \in [0,1]$, the code of~\cite{florian} allows for a general
coordinate transformation of the form $r = r(x)$, so that
either compactified - with future null infinity being the outermost
radial grid point - or non-compactified grids can be used.
This will allow for the unambiguous computation of the gravitational
radiation at ${\cal I}^+$ in our planned gravitational core collapse
simulations, by simply reading off the news function at the outermost
radial grid point. The hydrodynamic equations are formulated as 
in~\cite{pf2000} and solved using HRSC schemes.

As a simplified model for a self-gravitating relativistic
star~\cite{florian} have considered the spherically symmetric
solution of the general relativistic hydrostatic equation, the
so-called Tolman-Oppenheimer-Volkoff equation, with a polytropic
EoS $p = K \rho^{\Gamma}$. Equilibrium models were used to
check the long-term stability of the code and its convergence
properties. The code has proven to be stable for evolution times
much longer than the characteristic light-crossing times of the
different models considered. Spherically symmetric simulations
have also shown the expected second order accuracy of the code.

Following~\cite{font5} we have also checked the code on a dynamical
evolution of an unstable spherical initial model. In such a model the sign of
the truncation error of the numerical scheme controls the fate of the
evolution. In the code this sign is such that the unstable star
``migrates" to the stable branch of the sequence of equilibrium
models. In such a situation, the rest-mass of the star has to be
conserved throughout the migration. Despite the fact that this
mechanism cannot occur in nature - unstable stars can only collapse
to more compact configurations - and as such it is an academic problem,
it represents, nevertheless, an important test of the accuracy and 
self-consistency of the code in a highly dynamical situation.

\begin{figure}[t]
\begin{center}
\includegraphics[width=.85\textwidth,height=0.5\textwidth]
{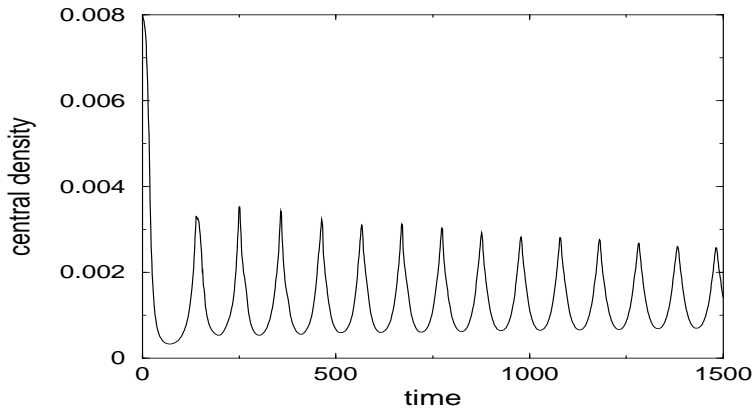}
\caption{Evolution of the central rest-mass
density during the migration of an unstable
relativistic star ($N=1, K=100, M=1.447 M_{\odot}, \rho_c=8.0 \times
10^{-3}; G=c=M_{\odot}=1$) to a stable model with the same
rest-mass. The central density of the (final) stable configuration
is $\rho_c=1.35\times 10^{-3}$. The evolution shows the expected
behavior. Since we are using a polytropic EoS, the amplitude of
the oscillations is essentially undamped for the evolution times
shown.}
\label{migration}
\end{center}
\end{figure}

As in~\cite{font5} we have constructed a $N=1$ ($\Gamma=1+1/N=2$),
$K=100$ polytropic star with mass $M=1.447\;M_\odot$ and central
rest-mass density $\rho_c=8.0\times 10^{-3}$ (in units in
which $G=c=M_{\odot}=1$).
Fig.~\ref{migration} shows the evolution of the
central density up to a final time of $u=1500$. On a
very short dynamical timescale the star rapidly expands and its central
rest-mass density drops well below its initial value, less than
$\rho_c=1.35\times 10^{-3}$, the central rest-mass density of
the stable model of the same rest-mass.
During the rapid decrease of the central density, the star
acquires a large radial momentum. The star then enters a phase of
large amplitude radial oscillations around the stable equilibrium
model. As Fig.~\ref{migration} shows the code is
able to accurately recover (asymptotically)
the expected values of the stable model. Furthermore, its evolution
is completely similar to that obtained with an independent fully
three-dimensional code in Cartesian coordinates~\cite{font5}.

The evolution shown in Fig.~\ref{migration} allows to study large
amplitude oscillations of relativistic stars, which cannot be
treated accurately by linear perturbation theory. These oscillations
could occur after a supernova core-collapse~\cite{dimmelmeier}
or after an accretion-induced collapse of a white dwarf.

To end this section we briefly discuss
a global energy conservation test of the axisymmetric characteristic
code which was presented in~\cite{florian}. For that purpose Siebel et
al~\cite{florian} used a strong ingoing gravitational wave to
perturb an equilibrium relativistic star:
\begin{equation}
\hat{\gamma}\equiv\frac{\gamma}{\sin\theta^2}
= 0.05 \ e^{-2 (r-4)^{2}} \ e^{-4 y^{2}}.
\end{equation}

\begin{figure}[t]
\begin{center}
\includegraphics[width=.85\textwidth,height=0.5\textwidth]{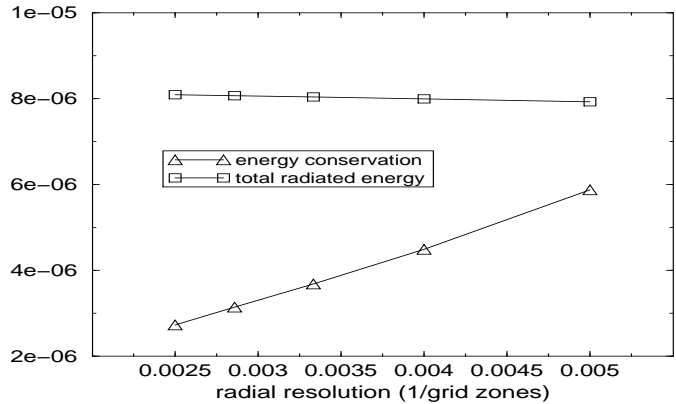}
\caption{Convergence properties of a global energy conservation test.
The difference between the initial and final Bondi mass converges in an
almost second order way to the value given by the energy which is carried away
by gravitational waves. The figure is taken from~\cite{florian}.}
\label{ec}
\end{center}
\end{figure}

The (nonlinear) initial pulse induces large velocities in the
fluid of the star which give rise to ``strong" outgoing gravitational
waves, i.e. with an energy larger than the numerical errors involved
in the calculation of the Bondi mass for a given resolution
and integration time. If $M$ is the Bondi mass and $P$ is the total
energy radiated away by gravitational waves at future null infinity
${\cal I}^+$, the convergence of the
quantity
\begin{equation}
ec := M|_{u=0} - M|_{u=u_{*}>0} - P|_{[0,u_{*}]}
\end{equation}
to zero represents a very severe global test of the numerical code.
Satisfactory convergence results under grid refinement are shown in 
Fig.~\ref{ec}.

\section{Summary}

The article has dealt with presenting some concepts and
applications in relativistic astrophysics of a particular 
class of finite difference
numerical schemes based on Riemann solvers, specifically
designed for nonlinear hyperbolic systems of conservation
laws.

Such schemes have been discussed in the context of the general
relativistic hydrodynamic equations. Nevertheless, the algorithms
presented are general enough to be applicable to other hyperbolic
systems such as the Einstein equations (when appropriately
formulated, as e.g. in Friedrich's conformal
approach~\cite{friedrich2}; see also~\cite{reula}). While
this may not be strictly necessary for vacuum spacetimes, it may
become relevant when dealing with nonvanishing stress-energy
tensors.

The use of conservative algorithms based upon the characteristic
structure of the hydrodynamic equations, developed during the
last decade building on ideas first applied in Newtonian
hydrodynamics, provides a robust methodology to obtain stable
and accurate solutions even in the presence of discontinuities.
This has become apparent since the early 1990s~\cite{mim}.

The knowledge of the wave structure of the equations is the
essential building block in the construction of the so-called
linearized Riemann solvers. The increasing use of these solvers
in relativistic hydrodynamics has proved successful in handling
complex flows, with high Lorentz factors and strong shocks,
superseding more traditional methods based on artificial 
viscosity~\cite{ibanez}.

In the last part of the article we have discussed some astrophysical
applications of such schemes, using the coupled system of the 
(characteristic) Einstein and hydrodynamic equations. Examples
involving the gravitational collapse of supermassive stars and
the evolution of relativistic compact stars have been presented.

\section*{Acknowledgements}
I am grateful to J\"org Frauendiener and Helmut Friedrich for their
invitation to participate in the workshop and to contribute to 
the proceedings. The work presented here is the result of
collaborations with a number of colleagues, in particular,
Jos\'e Mar\'{\i}a Ib\'a\~nez, Thomas Janka, Felix Linke,
Chema Mart\'{\i}, Ewald M\"uller, Philippos Papadopoulos and 
Florian Siebel. I would like to thank especially Chema
Mart\'{\i}, Ewald M\"uller and Florian Siebel for carefully 
reading the manuscript.

%

\end{document}